\documentclass[aps,prl,preprint,showpacs,preprintnumbers,amsmath,amssymb,groupedaddress]{revtex4-1}

\usepackage{soul}

\usepackage[dvipdfmx]{graphicx}
\usepackage{dcolumn}
\usepackage{bm}
\usepackage{color}
\definecolor{Red}  {rgb}{1,0,0}
\definecolor{Green}{rgb}{0,1,0}
\definecolor{Blue} {rgb}{0,0,1}
\newcommand {\delspan}[1] {}
\newcommand {\addspan}[1] {#1}

\newcommand {\bfv}[1] {{\boldsymbol {#1}}}

\newcommand {\IND} {\ }

\newcommand\Rey{\mbox{\textit{Re}}}  
\newcommand\removed[1]{}

\newcommand\SEC[1]{}

\usepackage{ulem}

\begin{document}

\title{Bifurcation aspect of wide-gap spherical Couette flow emphasizing polygonal coherence and wave numbers observed over transitional Reynolds numbers}


\author{}
\affiliation{}

\author{Fumitoshi Goto}
\affiliation{
  Department of Pure and Applied Physics,
  Faculty of Engineering Science, Kansai University, Osaka, 564-8680, Japan
}
\author{Tomoaki Itano}
\email{itano@kansai-u.ac.jp}
\affiliation{
  Department of Pure and Applied Physics,
  Faculty of Engineering Science, Kansai University, Osaka, 564-8680, Japan
}
\author{Masako Sugihara-Seki}
\affiliation{
  Department of Pure and Applied Physics,
  Faculty of Engineering Science, Kansai University, Osaka, 564-8680, Japan
}
\author{Takahiro Adachi}
\affiliation{
  Department of Systems Design Engineering,
  Akita University, Akita, 010-8502, Japan
}


\date{\today}

\begin{abstract}
  This study numerically investigates the bifurcation aspect of the wide-gap spherical Couette flow (SCF), with an emphasis on the competition among polygonal coherence with different wave numbers observed over transitional Reynolds numbers.
  Focusing on a representative case, the half-radius ratio $\eta=1/2$, we confirm that the axisymmetric state becomes unstable over the first transitional Reynolds number at which the 4-fold spiral state bifurcates, using the continuation method based on the Newton--Raphson algorithm. The Galerkin-spectral method was employed to numerically solve the governing equations.
  It is found that the 3-fold spiral state bifurcates from the axisymmetric state at a slightly higher Reynolds number than the first transitional Reynolds number.
  The attraction of the 3-fold spiral state expands rapidly with an increase in the Reynolds number, which is determined by verifying the distance of the unstable periodic-like state to both spiral states in the state space.
  This aspect of the state space explains the experimentally bistable realization of different equilibrium states over the first transitional Reynolds number.
  This study also found that the periodic-like state is composed of the $3$- and $4$-fold spiral states, similar to a beat with two different frequencies.
  
\end{abstract}



\maketitle

%
%
%
%

\section{Introduction}

\IND
An incompressible Newtonian fluid confined between concentric spherical boundaries rotating differentially, namely the {\it spherical Couette flow} (SCF), is one of the canonical flows.
The SCF is a configuration applicable to astrophysical and geophysical flows in the planetary core, atmosphere, and ocean.
Stimulated by geophysical issues, experimental studies on the SCF were initiated by a few pioneers during the last quarter of the past century \cite{Cha61,Bus75,Zeb83,Kid97,Sak99,Fow04,Feu11}.
In particular, the case involving a rotating inner sphere and stationary outer sphere has been largely investigated; this case is controlled by the Reynolds number conventionally defined as $\displaystyle \Rey=r_{\rm in}^2\Omega_{\rm in}/\nu$.
Based on early investigations with several different combinations of spherical boundaries and various radii, it has been pointed out that SCF is fairly complicated compared to the cylindrical Taylor--Couette flow, despite the apparent similarity \cite{Mun75,Bel84}.
This difference is likely attributed to the presence of the meridional circulation and inflection point in the zonal profile in the hemisphere, which become more influential in the case of the larger gap between boundaries, that is, a small radius ratio.
The case of a small radius ratio $\eta\le 3/4$, which has been conventionally named ``{\it wide-gap}" or "{\it thick layer}'', exhibits somewhat puzzling sequential transitions to turbulence involving non-uniqueness and hysteresis.

\IND
The first transition in a wide-gap SCF is triggered by the polygonal traveling waves, which are similar to the flow between two rotating planar disks.
In general, a rotating shear flow always favors polygonal coherence around the poles. Regular polygonal patterns in the polar jet streams on Jupiter and Saturn were observed by recent spacecraft missions, which are relevant to the polygonal coherence in an SCF \cite{God88}.
The observed polygonal waves in a wide-gap SCF, which form a sinuous disturbance at the mid-latitudes propagating at a significantly low angular phase velocity in the zonal direction, were visualized as a spiral pattern with $m$ equally spaced arms extending from the poles to the equatorial zone in each hemisphere.
The transitional Reynolds numbers reported in a previous experimental study were $\Rey_{\rm cr}=2628$ ($m=6$) at $\eta=3/4$, $\Rey_{\rm cr}=1244$ ($m=5$) at $\eta=2/3$, and $\Rey_{\rm cr}=489$ ($m=4$) at $\eta=1/2$, where $m$ was the zonal wave number at the critical Reynolds number \cite{Jun00}.
It was confirmed that some of these values align remarkably well with theoretical predictions based on linear stability analysis \cite{Dum94,Ara97,Hol06}.
A recent investigation with a further increase in the Reynolds number suggested relaminarization, non-uniqueness, and hysteresis in successive transition processes in an SCF \cite{Egb95,Wul99,Nak02,Abb18a}.
These aspects in a wide-gap SCF can be regarded as a type of competition among different polygonal structures, which plays an important role in the route to turbulence.
For example, Wulf et al. investigated successive transition processes undergoing several mode changes and concluded that the Ruelle-Takens-Newhouse scenario, associated with a few polygonal modes, represented a route to turbulence in a wide-gap SCF \cite{Wul99}.
The variety of routes to turbulence still remains an open question in SCF.

\IND
Here, we focus on a representative but insufficiently investigated radius ratio, $\eta=r_{\rm in}/r_{\rm out}=1/2$.
Belyaef et al. estimated the first transitional Reynolds number $\Rey_{\rm cr}=460$ at $\eta=1/2$ based on experimental studies of the power spectra of an SCF using laser Doppler velocimetry \cite{Bel84}.
The emerged flow consisted of four vortices ($m_{\rm cr}=4$) in every hemisphere, displaced in staggered rows with respect to the equator.
By quasi-statically increasing $\Rey$ further, a new wave regime emerged at the second transitional Reynolds number characterized by the $m=3$ regime instead of the $m=4$ regime, via a period doubling bifurcation of $m=4$ with a sub-harmonic frequency.
However, the $m=3$ regime attained at a higher Reynolds number can be quasi-statically maintained near the first transitional Reynolds number with reorganization of the $m=4$ regime.
This hysteresis was observed in the relatively low Reynolds number range, as previously reported by Belyaef and Yavorskaya \cite{Bel91}.
They also suggested that a decrease in $\Rey$ after an abrupt change in $\Rey$ from $\Rey<550$ to $\Rey>660$ creates a temporal chaotic regime.
The further decrease in $\Rey$ from the chaotic state leads to an irreversible transition for a spatial structure with $m=3$ (see Fig.1 in Ref. \cite{Bel91}).
In an overview of experimental and numerical studies, including the case of $\eta=1/2$, Junk and Egbers concluded that the transitional Reynolds number was 489 and pointed out that similar flow regimes with a small number of spiral waves were observed before the flow became turbulent at a higher Reynolds number \cite{Jun00}.


\SEC{Motivation}
The primary motivation of the current work is to study the exchange of the stability of the $m=3$ and $m=4$ regimes at a representative radius ratio $\eta=1/2$, which has been solved insufficiently.
The aforementioned experimental observation implies that the $m=4$ regime originates in the axisymmetric state and becomes stable over the first transitional Reynolds number, where the axisymmetric state is unstable.
However, it is plausible that the $m=3$ regime does not originate in the $m=4$ regime and instead bifurcates from the unstable axisymmetric state.
From the viewpoint of the deterministic dynamical system, it is significant how the $m=3$ regime is established as a stable state over the second critical Reynolds number.
Through the present study, we will elucidate the formation of the $m=3$ regime in the state space, which has remained unclear thus far.

\SEC{Organization of this paper}
The remainder of this paper is organized as follows.
In the next section, we briefly describe the nondimensionalized governing equations and numerical setup of our system.
In Section 3, we present the numerically obtained non-axisymmetric states  and discuss the morphology of the vortex in comparison with previous studies.
In the latter part of the section, the $m=4$ and $m=3$ states are explored using the Newton--Raphson algorithm, and the first and second transitional Reynolds numbers are specified.
We numerically confirm the bistability of these regimes and discuss the basin of attraction of non-axisymmetric states.
The paper concludes with brief remarks on the unstable saddle-like state between these states.




\section{Formulation}
\begin{figure}[h]
  \centering
  \includegraphics[angle=0,width=0.75\columnwidth]{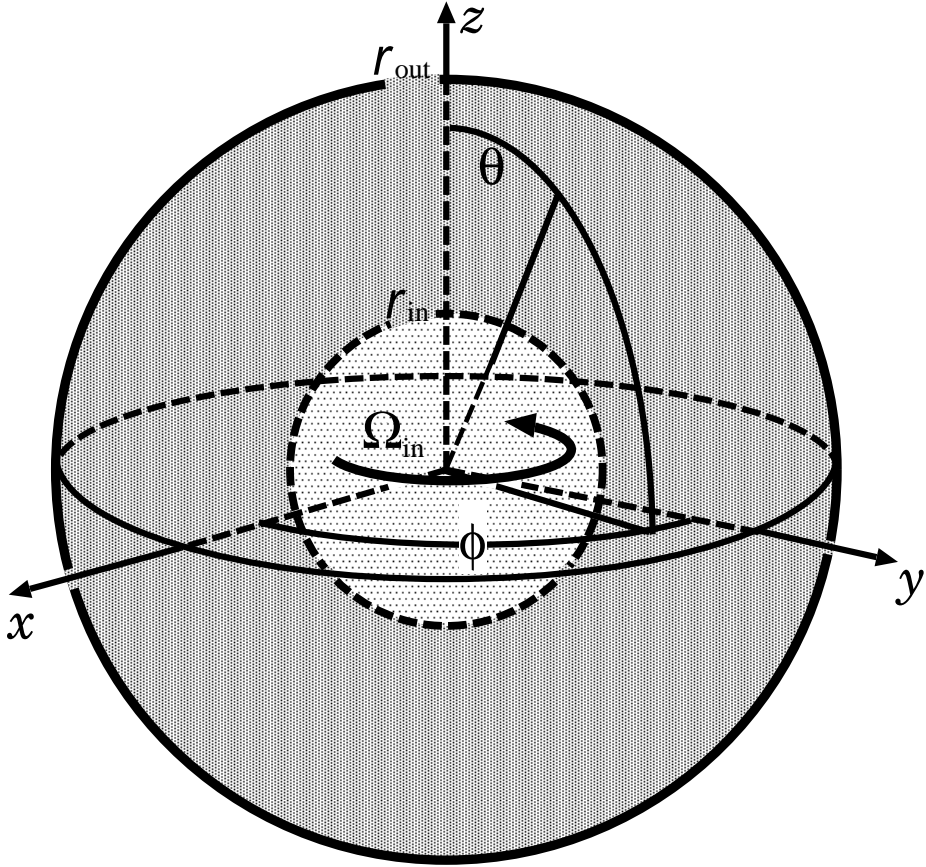}
  \caption{
    Configuration of the present study. 
}
  \label{fig.config}
\end{figure}

\SEC{Direct Numerical Simulation}
\IND
The spherical coordinate represented by the radial, zenith, and zonal components, $(r,\vartheta, \varphi)$, is centered at the origin.
The incompressible fluid is confined between the inner and outer spheres with radii $r_{\rm in}$ and $r_{\rm out}$, respectively, as shown in Fig. \ref{fig.config}. The inner sphere rotates at a constant angular velocity $\Omega_{\rm in}$ with respect to the $z$ axis.
    The nondimensional geometrical parameter is either the gap ratio $\beta=(r_{\rm out}-r_{\rm in})/r_{\rm in}$ or aspect ratio $\eta=r_{\rm in}/r_{\rm out}$.

The Galerkin-spectral method was employed to numerically solve the governing equations, 
\[ \frac{\partial\bfv{u}}{\partial t} + (\bfv{u}\cdot\bfv{\nabla})\bfv{u}=-\frac{1}{\rho_0}\bfv{\nabla} p+\nu_0 \bfv{\nabla}^2\bfv{u}\]
under the incompressible condition $\bfv{\nabla}\cdot\bfv{u}=0$.
Due to the divergence-free constraint, toroidal and poloidal decomposition were invoked with regard to the radial direction, $\bfv{u}(\bfv{r})=u_{\rm 0}(r,\theta)\bfv{e}_\phi(r,\theta)+\bfv{\nabla}\times\Bigl\{-\bfv{r}\Psi(\bfv{r})+\bfv{\nabla}\times\bigl(\bfv{r}\Phi(\bfv{r})\bigr)\Bigr\}$, where the magnitude of the Stokesian shear flow $u_{\rm 0}(r,\theta)$ is proportional to $\Omega_{\rm in}$ and satisfies $\bfv{\nabla}^2(u_0\bfv{e}_\phi)=0$, boundary conditions, $u_0(r_{\rm in})=r_{\rm in}\Omega_{\rm in}$, and $u_0(r_{\rm out})=0$ \cite{Lan87}.
Hereafter, the governing equation is nondimensionalized by the gap half-width $\Delta r=(r_{\rm out}-r_{\rm in})/2$ and the viscous diffusion time $\Delta r^2/\nu_0$. Therefore, the system is uniquely determined only by the geometrical parameter $\eta=r_{\rm in}/r_{\rm out}$ and Reynolds number $\displaystyle \Rey$. 
The toroidal and poloidal components are spatially expanded into a series of spherical harmonics $Y^m_l(\theta,\phi)$ with a polar wave number $l$ and zonal wave number $m$ through the aid of numerical libraries on spherical harmonics utilizing the Gauss-Lobatto collocation method \cite{Sch13,Fri05}.
Adapting the second-order Adams--Bashforth method, complemented by Crank--Nikolson for temporal discretization, we converted the nondimensionalized governing equation to an equivalent inhomogeneous Helmholtz equation.
This Helmholtz equation is equivalent to a set of linear algebraic equations for the expansion coefficients $\Psi_{l,m,n}$ and $\Phi_{l,m,n}$ with the expansion into a series of modified Chebyshev polynomials $(1-y^2)^k T_n(y)$ ($k=1$ for $\Psi$ and $k=2$ for $\Phi$), where $y=\bigl(r-r_0\bigr)/{\Delta r}$ and $\displaystyle r_0=\frac{r_{\rm in}+r_{\rm out}}{2}$. This was also used in our previous study \cite{Ita09}.
This set of equations can be solved numerically using the LAPACK libraries \cite{And99}.
Unless noted, we used the truncation levels $(l_{\rm max},m_{\rm max},n_{\rm max})=(30,30,32)$.

\SEC{Basic laminar state}
\IND
The developed numerical code \cite{Ina19} was validated for ${\Rey}<600$ using a quantitative comparison with previous experimental and numerical results \cite{Hol06,Nak02,Wul99}.
Numerical integration was used to obtain some equilibrium states at $\eta=1/2$ around the transitional Reynolds numbers, beginning initially with a Stokesian flow with a small disturbance that was artificially generated by a series of random numbers.
The basic laminar flow preferred under the first transitional ${\Rey}$ consisted of the Stokesian shear flow perturbed by two zonal momentum cells in the northern and southern hemispheres. These were divided by a strong and local radial outward flow developing at the equatorial zone via the inertial (centrifugal) force of the rotating inner sphere.
The basic laminar flow satisfied the axisymmetry $\frac{\partial}{\partial\phi}\Phi=\frac{\partial}{\partial\phi}\Psi=0$ and equatorial symmetry $\bigl(\Phi(\pi-\theta),\Psi(\pi-\theta)\bigr)=\bigl(\Phi(\theta),-\Psi(\theta)\bigr)$.
A fluid element initially confined in a hemisphere travels from the equator to the pole along the outer spherical boundary and then back to the equator along the rotating inner boundary.
It was quantitatively confirmed that the magnitudes of the $u_r$ and $u_\phi$ components of the numerically obtained axisymmetric flow aligned with those in a previous numerical study, with an error of a few percentage points \cite{Hol06}.
As Hollerbach {\it et al.} reported, the magnitude of the radial component obtained numerically is comparable to that of the zonal component near the equator, which is characteristic of a wide-gap SCF but not narrow-gap cases \cite{Hol06}.

\section{Results}
\begin{figure}
  \centerline{
    \includegraphics[angle=0,width=1.10\columnwidth]{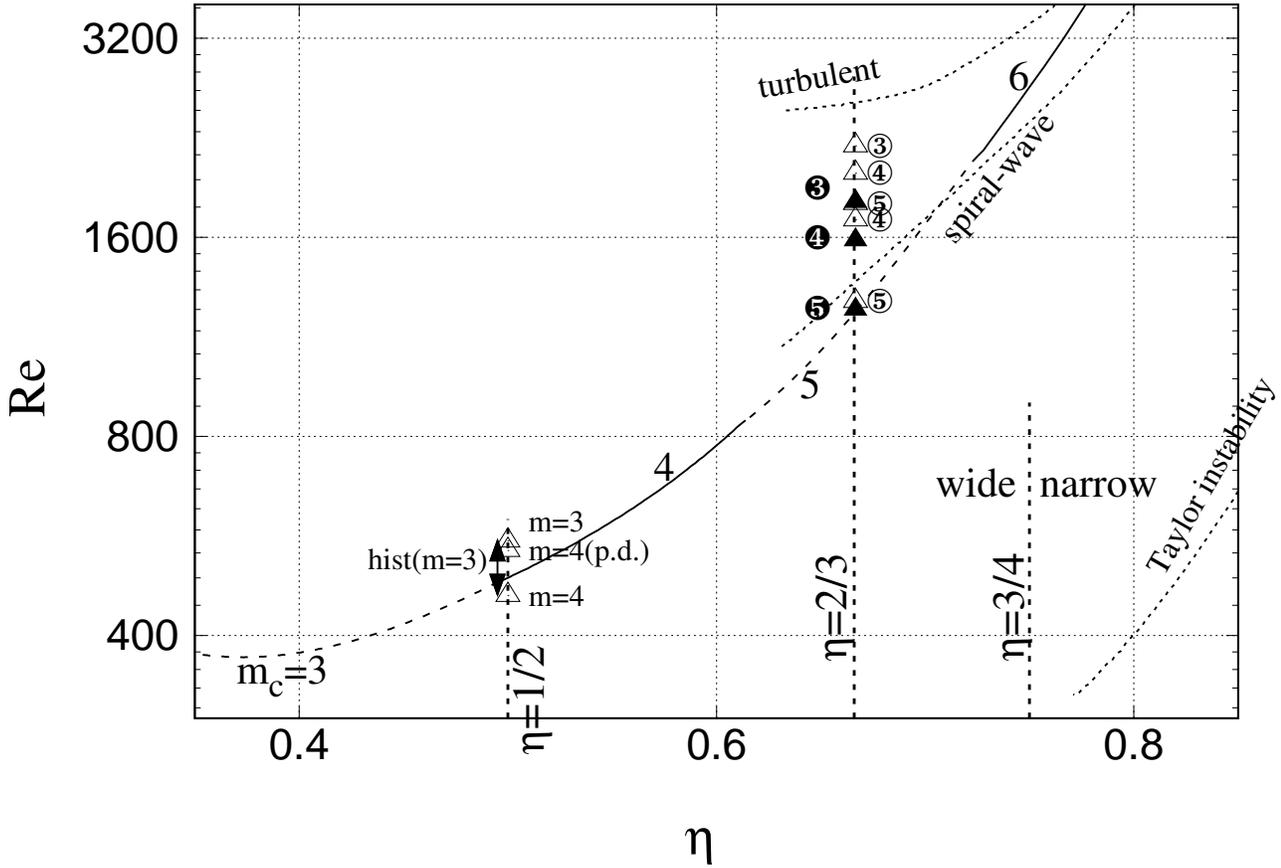} }
  \caption{
    SCF experiences the first transition on the route to turbulence in two different manners depending on the value of the geometrical parameter, $\eta$. The solid and dashed curves are the neutral curve of the axisymmetric state against an infinitesimal perturbation of the indexed even and odd zonal wave numbers, respectively. The numbers in the circles and triangles indicate the wave numbers of the spiral state observed in previous studies.
      }
  \label{fig.eta_vs_Re}
\end{figure}

\SEC{Secondary wave}
\IND
With an increase in the Reynolds number, the SCF experiences the first transition on the route to turbulence in two different manners depending on the value of the geometrical parameter, as shown in Fig. \ref{fig.eta_vs_Re}. In the narrow-gap cases ($\eta>3/4$), the transition is primarily initiated by axisymmetric Taylor vortices with successive instability, as observed in the cylindrical Taylor--Couette flow.
    Conversely, in the wide-gap cases ($\eta < 3/4$), the first transition is triggered by non-axisymmetric secondary waves extending from the poles.
    The solid and dashed curves, which were reconstructed based on values read from the graph in Ref. \cite{Hol06}, are the neutral curves of the axisymmetric state against an infinitesimal perturbation of the indexed even and odd zonal wave numbers, respectively.
    The number of waves in a hemisphere at the transition is determined by the value of $\eta$ and can be calculated by the linear stability analysis of the axisymmetric state.
    The numbers in the circles and triangles indicate the wave numbers of the spiral state observed in previous studies (Ref. \cite{Bel84} at $\eta=1/2$ and Ref. \cite{Egb95,Abb18a} at $\eta=2/3$), and they are located at the values $(\eta,Ra)$ where the state was obtained experimentally or numerically.

Egbers and Rath employed experimental studies for several values of $\eta$ in ``{\it narrow-gap}'' cases and found that the transition was initiated by the axisymmetric Taylor vortices with successive instability where $\eta > 3/4$, as observed in the cylindrical Taylor--Couette flow \cite{Egb95}.
However, in ``{\it wide-gap}'' cases, the first transition was triggered in the absence of Taylor vortices by the non-axisymmetric secondary waves extending from the poles where $\eta < 3/4$.
The flow structure of the secondary wave consisted of a set of sinuous waves at the equator traveling in the zonal direction, which connected with the spiral disturbance extending in each hemisphere from the poles.
Hereafter, we refer to the secondary wave with wave number $m$, which is the number of spiral waves extending from a pole, as an {\it $m$-fold spiral state}.
The dependencies of the transitional Reynolds number on the value of $\eta$ are listed in Ref. \cite{Jun00}.
The wave number $m$ selected at the first transition depends on the value of $\eta$.
In the wide-gap case, the larger the value of $\eta$, the larger is the wave number.

\IND
Moreover, an experimental study conducted considering $\eta=2/3$ reported that a further increase in the Reynolds number caused a successive sequence of transitions with a decrease in the wave number \cite{Egb95}. In other words, a spiral state with a smaller wave number was dominant as the Reynolds number increased.
They reported that $m=5$, $m=4$, and $m=3$ were observed at $\Rey=1250$, $\Rey=1600$, and $\Rey=1900$, respectively. Clearly, the system exhibited a somewhat complicated hysteresis at these successive transitional Reynolds numbers. This hysteresis has also been reported in a numerical study considering $\eta=2/3$, where the $m=5$-fold spiral state was obtained at $\Rey=1280$, $m=4$ at $\Rey=1700$, $m=5$ at $\Rey=1800$, $m=4$ at $\Rey=2000$, and $m=3$ at $\Rey=2200$ \cite{Abb18a}. This is represented by the numbers marked with circles at $\eta=2/3$ in Fig. \ref{fig.eta_vs_Re}.
The critical wave number at the first transitional Reynolds number was determined by $\eta$, whereas the ratio of the representative magnitude of the meridional circulation to the rotational speed of the inner sphere affected a successive sequence of transitions \cite{Ina19}.
The increase in $\Rey$ primarily enhanced the meridional circulation, whereas mixing in the gap induced a non-axisymmetric flow, weakening the circulation.
Once $\Rey$ exceeds a critical value, a further increase in $\Rey$ enhancing the mixing reduces the meridional circulation, which may lead to a spiral state with a reduced wave number.
The decrease in the wave number of the non-axisymmetric flow may be attributed to the reduction in the ratio of the meridional circulation to the inner sphere rotational speed.

\SEC{Bistable system}
Previous experimental and numerical studies have shown that hysteresis over the transitional Reynolds number is common in wide-gap SCFs.
This implies that our system with $\eta=1/2$ is a bistable system, where different stable spiral states are attained from different initial states.
Based on an experimental study of the appearance of a single frequency in the power spectra, Belyaef et al. estimated the first transitional Reynolds number at $\eta=1/2$ to be $460$ with $m=4$ \cite{Bel84}.
They reported that an equilibrium state characterized by $m=3$ vortices was realized over $\Rey=556$ instead of $m=4$ vortices, via a period doubling bifurcation excited by a sub-harmonic frequency over $\Rey=538$.
Moreover, they studied the lower $\Rey$ limit that the $m=3$ vortex regime survives by decreasing $\Rey$ from $\Rey=556$.
This regime was not realized until the $m=4$ fold state reappeared at $\Rey=462$.
This suggests that our system with $\eta=1/2$ is also a bistable system over a certain second transitional Reynolds number.
The first transitional Reynolds number at $\eta=1/2$ was solved as $\Rey=489$ in later studies \cite{Dum94,Jun00}.
Additionally, it should be noted that the wave number selected at the first transition in the case of $\eta=1/2$ is $m=4$, whereas the first transition at $\eta \lesssim 0.5$ is triggered by the $m=3$-fold spiral state \cite{Hol06}.
This may be relevant to the competition between the $m=3$- and $m=4$- fold spiral states near the first transition at $\eta=1/2$.

\begin{figure}
    \includegraphics[angle=0,width=0.98\columnwidth]{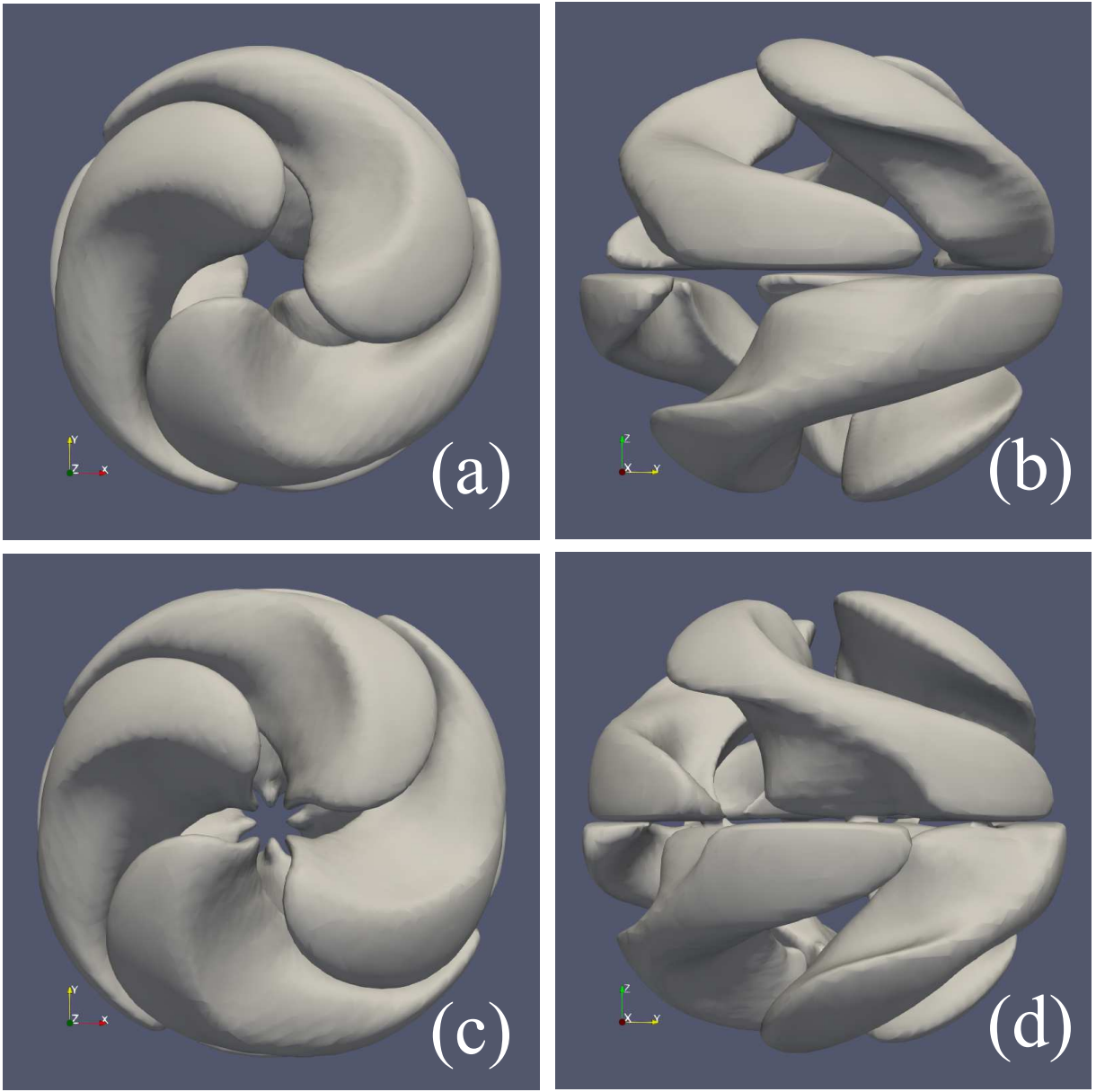}
  \caption{
    Views of (a)Polar 3-fold, (b) equatorial 3-fold, (c) polar 4-fold, and (d) equatorial 4-fold. 
  }
  \label{snapshots}
\end{figure}

\SEC{Newton-Raphson method}
First, we carried out numerical trial computations at $\eta=1/2$ by initially reducing all modes in the expansion coefficients of $\Phi$ and $\Psi$, except for $m=0,3,4$.
The computations at $\Rey=560$ approximately settled in either the $m=3$ or $m=4$-fold spiral states, depending on the disturbance in the initial condition.
This implies that the $3$- and $4$-fold spiral states are stable attractors and that the basic state is a repellor.
The obtained $m$-fold spiral states spontaneously satisfy $\Phi(r,\theta,\phi+2\pi/m)=\Phi(r,\theta,\phi)$ and $\Psi(r,\theta,\phi+2\pi/m)=\Psi(r,\theta,\phi)$.
In addition, the flow pattern appearing in both hemispheres shifted toward each other by half the wavelength, such that the reflection symmetry to the equatorial plane, $\Phi(r,\pi-\theta,\phi)=-\Phi(r,\theta,\phi)$ and $\Psi(r,\pi-\theta,\phi)=\Psi(r,\theta,\phi)$ were satisfied.
Taking into account these symmetries and the assumption of a rotating wave solution with a constant angular velocity, $\omega_\phi$, we substituted $\frac{\partial}{\partial t}=-\omega_\phi \frac{\partial}{\partial \phi}$ in the original governing equation, where the solution was assumed to constantly rotate around the $z$ axis along the positive $\phi$ direction.
A Galerkin-type reduced quadratic equation was deduced from the substitution for the expansion coefficients $\Psi_{l,m,n}$ and $\Phi_{l,m,n}$.
Here, the freedom of $\omega_\phi$ for the $m$-fold spiral state could be specified by fixing the zonal phase of the state.
We subsequently solved the quadratic equations using the iterative Newton--Raphson method with the aid of LAPACK libraries \cite{And99}.
The polar and equatorial views of the 3- and 4-fold spiral states were converged at $(\eta,\Rey) = (1/2, 490)$ and are visualized in Fig. \ref{snapshots} by an isosurface of the $\Phi$ deviation from the mean value. 
    These states satisfy not only the $m$-fold symmetry with respect to a pole but also a reflection symmetry with respect to the equatorial plane.
Note that the time constant of the least stable mode of the spiral states tends to almost vanish, particularly near the transitional Reynolds numbers.
The Newton--Raphson algorithm has an advantage over the direct numerical simulation in that it is able to solve a less stable state. This enabled us to solve the phase angular velocity and obtain the states with a small time constant, which could be principally traceable using a long-term simulation.

\begin{figure}
  \centerline{
    \includegraphics[angle=0,width=1.25\columnwidth]{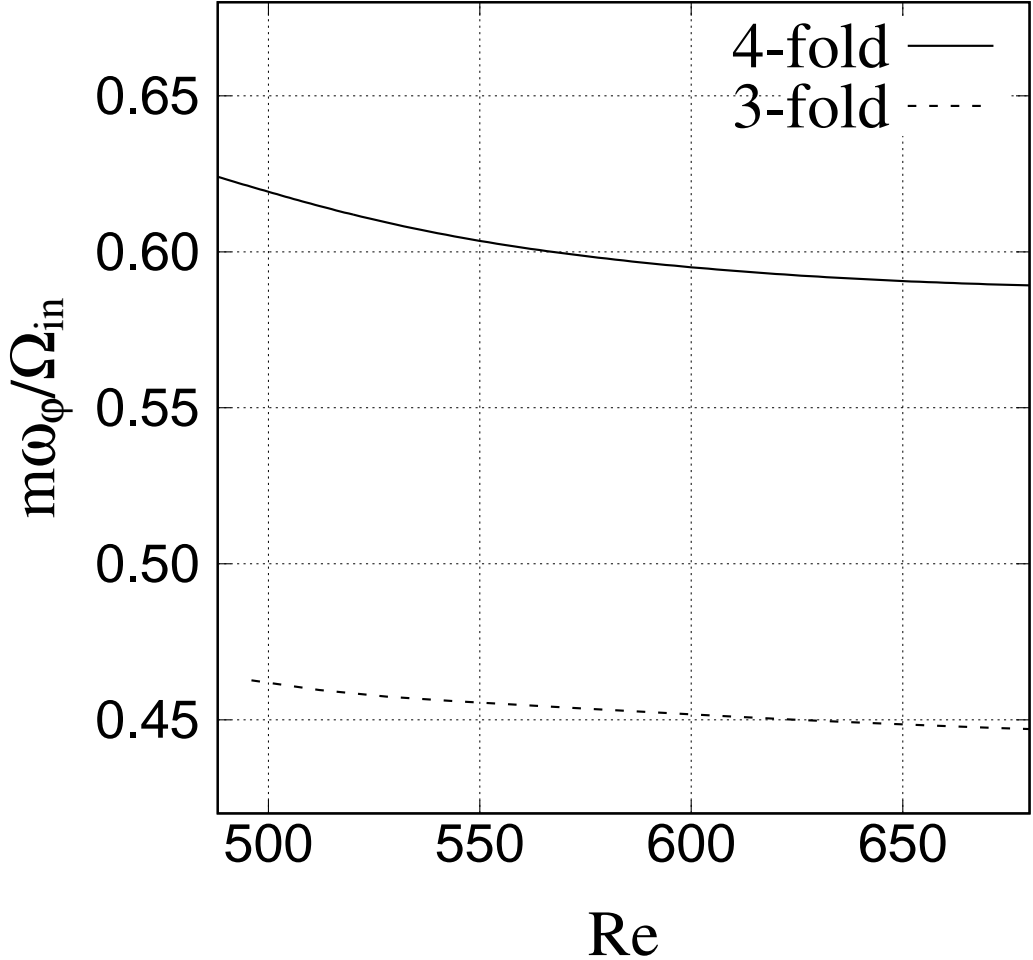}  }
  \caption{
    The dimensionless frequencies against $\Rey$, calculated from the phase angular velocities of $m=3$- and $m=4$-fold spiral states for $\eta=1/2$, decreased with an increase in $\Rey$.}
  \label{Newton omega}
\end{figure}

\SEC{Homotopy of $m$-fold spiral state}
We continuously obtained the $m=3$ and $m=4$-fold spiral states using the Newton–-Raphson algorithm by gradually varying $\Rey$ from the transitional Reynolds number up to $\Rey=800$.
For each $\Rey$, the phase angular velocity $\omega_\phi$ converged to a value through stepwise iterations of the method.
The dimensionless frequency, $m\omega_\phi/\Omega_{\rm in}$, is defined from the ratio of the phase angular velocity $\omega_\phi$ of the obtained state to the inner sphere angular velocity $\Omega_{\rm in}$; this was plotted against the Reynolds number, as shown in Fig. \ref{Newton omega}.
The frequencies for the $4$- and $3$-fold states at $\Rey=490$ and $\Rey=496$ are 0.623 and 0.463, respectively.
They gradually decreases with the increase of $\Rey$.
  Note that these coherent patterns rotate to the zonal direction with almost identical and relatively low angular velocities, approximately one sixth of the inner sphere rotation.
The experimentally obtained value of the dimensionless frequency would appear as sharp peaks in the power spectra if a pointwise velocity signal of the flow was measured experimentally by using laser Doppler velocimetry in order to monitor the transition to chaos by detecting the broadening of these spectral lines.
In particular, the frequency for the $4$-fold spiral state near the first transition was measured as 0.614 in a previous experiment, which is in reasonable agreement with the present numerical result \cite{Bel91}.
As shown in Fig. \ref{Newton omega}, the dimensionless frequency gradually decreases as the Reynolds number increases, which also resembles the behavior in Fig. 1, as reported by Belyaef and Yavorskaya \cite{Bel91}.

\begin{figure}
  \centerline{
    \includegraphics[angle=0,width=1.25\columnwidth]{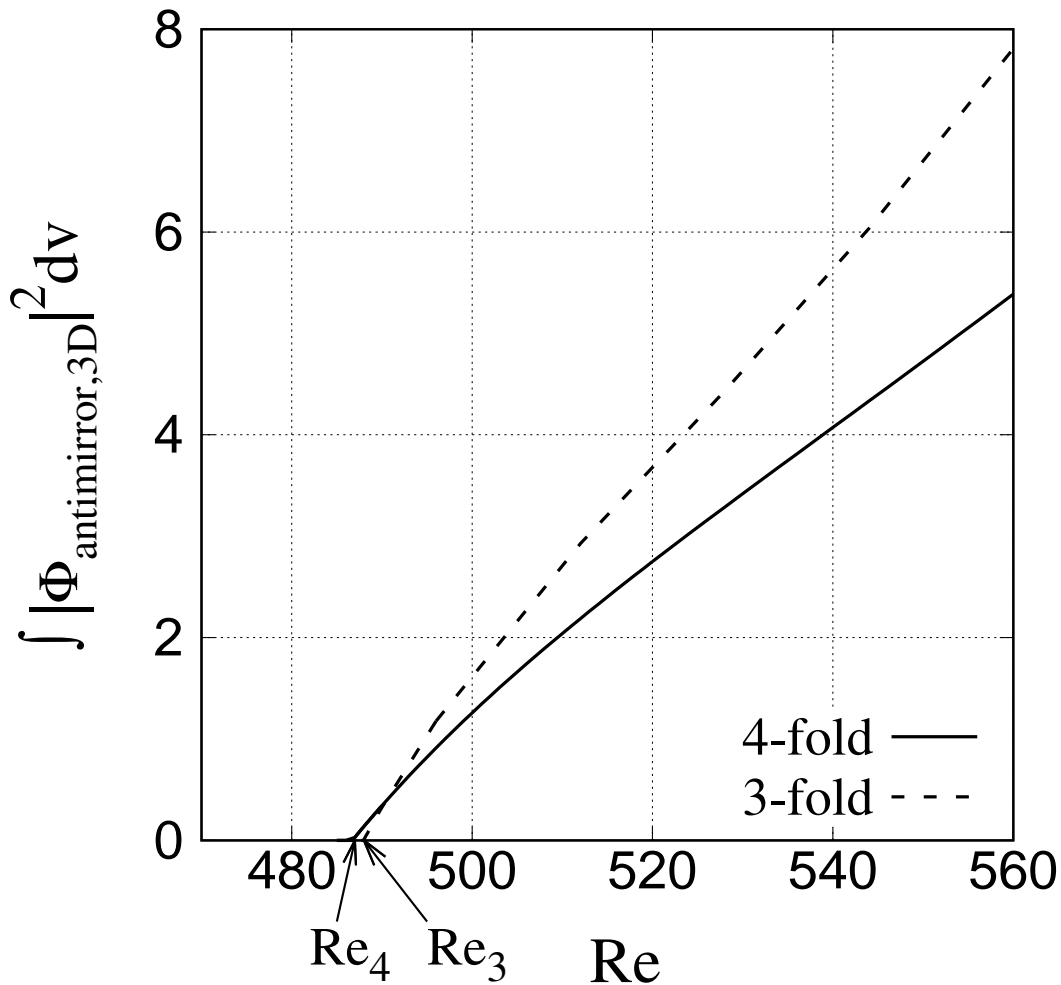}}
  \caption{Norm of the anti-mirror component of $Phi$ with $m\ne 0$ plotted against $\Rey$ for the $3$- and $4$-fold spiral states obtained at $\eta=1/2$.
    }
  \label{Newton}
\end{figure}

\SEC{Critical Reynolds number}
The transitional Reynolds number for $\eta=0.5$ was first specified as $\Rey_{\rm cr}=460$  by Belyaef et al. \cite{Bel84}. This was later numerically refined by Dumas \cite{Dum94} and Hollerbach \cite{Hol06}, and experimentally refined by Markus \& Egbers \cite{Jun00} to be $\Rey_{\rm cr}=489$.
The norm of the non-axisymmetric components of $\Phi$ for the $m=3$- and $m=4$-fold spiral states is defined as the integral of the squared velocity over the whole volume $\int |\Phi_{\rm antimirror,3D}| dv$, where $\Phi_{\rm antimirror,3D}=\frac{1}{2}\sum_{m\ne 0}\bigl(\Phi_m(r,\theta,\phi)-\Phi_m(r,\pi-\theta,\phi)\bigr)$, as shown in Fig. \ref{Newton}.
In the present calculation, the $3$- and $4$-fold spiral states converge to the axisymmetric states at $\Rey_4=486$ and $\Rey_3=487$, respectively. The $3$- and $4$-fold spiral states bifurcate from the axisymmetric state at different transitional Reynolds numbers, i.e., $\Rey=\Rey_{3}$ and $\Rey=\Rey_{4}$, respectively, where $\Rey_{4}<\Rey_{3}$.

The arithmetic discrepancy in the critical Reynolds number between the previous and present studies is attributed to the difference in the numerical truncation.
We confirmed that the $4$-fold spiral state converged to the axisymmetric state at $\Rey_4=489$ as the truncation level for the spherical polynomials increased from 30 to 60.

\section{Discussion}
\SEC{Basin boundary in reduced map}
The degree of $m$-fold spiral states can be measured by the magnitude of $\displaystyle X_{m}^2=\int dv |\Phi_{m}|^2$ and $\Phi_{m}(r,\theta,\phi)=\sum_{l} \Phi_{l,m}(r)Y_{l}^m(\theta,\phi)$. 
Let us project the full state space of our system into a reduced space spanned by $X_{3}$ and $X_{4}$.
Note that the time evolution of the state is illustrated as a trajectory in the projection and that the $3$- and $4$-fold spiral states correspond to fixed points on the abscissa and ordinate, respectively.
The basic state located at the origin of the map constitutes an unstable equilibrium over the first transitional Reynolds number.
We numerically confirmed in advance that the $3$- and $4$-fold states were stable and that the basic laminar state was unstable for Reynolds numbers in the present study.
Thus, both states were not only equilibrium states but also attractors in the state space, such that any infinitesimal perturbation against either equilibrium state decreased asymptotically.
Consequently, there must be at least one subspace (a super surface in the full state space) that separates the basins of these attractors.
We will hereafter refer to this as a {\it basin boundary} between the 3-  and 4-fold spiral states.

\SEC{Edge tracking}
Here, we will obtain the basin boundary in the map by employing an edge-tracking or shooting method, which was originally established as a tool to find an unstable steady solution embedded in a subcritical system, such as a planar Couette flow \cite{ita01,toh03,sch08,dug10}.
Hereafter, we denote the $m$-fold spiral states as $\bfv{x}_m$.
We used an intermediate state $\bfv{x}_{s}=(1-s)\bfv{x}_3+s \bfv{x}_4$ as the initial condition for a trial run, where $s$ was a parameter.
In the full state space, the continuous set of the initial condition $\bfv{x}_{s}$ for $0\le s\le 1$ corresponded to a line segment connecting the $m=3$ and $m=4$ spiral states.
The segment must intersect the basin boundary at least once because both equilibrium states at the ends of the line segment are stable.
We adjusted the value of $s$ such that the state approached neither the $m=3$ nor $m=4$ spiral states for as long as possible.
The edge-tracking indicated that the time evolution for $s<s_0$ asymptotically reached a 3-fold spiral state, while that for $s_0<s$ reached a 4-fold spiral state.
It is significant that the asymptotic state starting from $\bfv{X}_{s}$ is only classified by the magnitude of $s$.
 An infinite period is needed for the crossing point $s=s_0$ for the trial run $\bfv{x}(t)$ starting from $\bfv{x}_{s_0}$ to reach either attractor.
Principally, the trajectory starting with $s=s_0$ stays on the basin boundary, which is also a high-dimensional space.
Thus, the time variation of the state variables of the trajectory with $s_0$ could exhibit chaotic or turbulent behavior.

\begin{figure}
  \centerline{
    \includegraphics[angle=0,width=1.25\columnwidth]{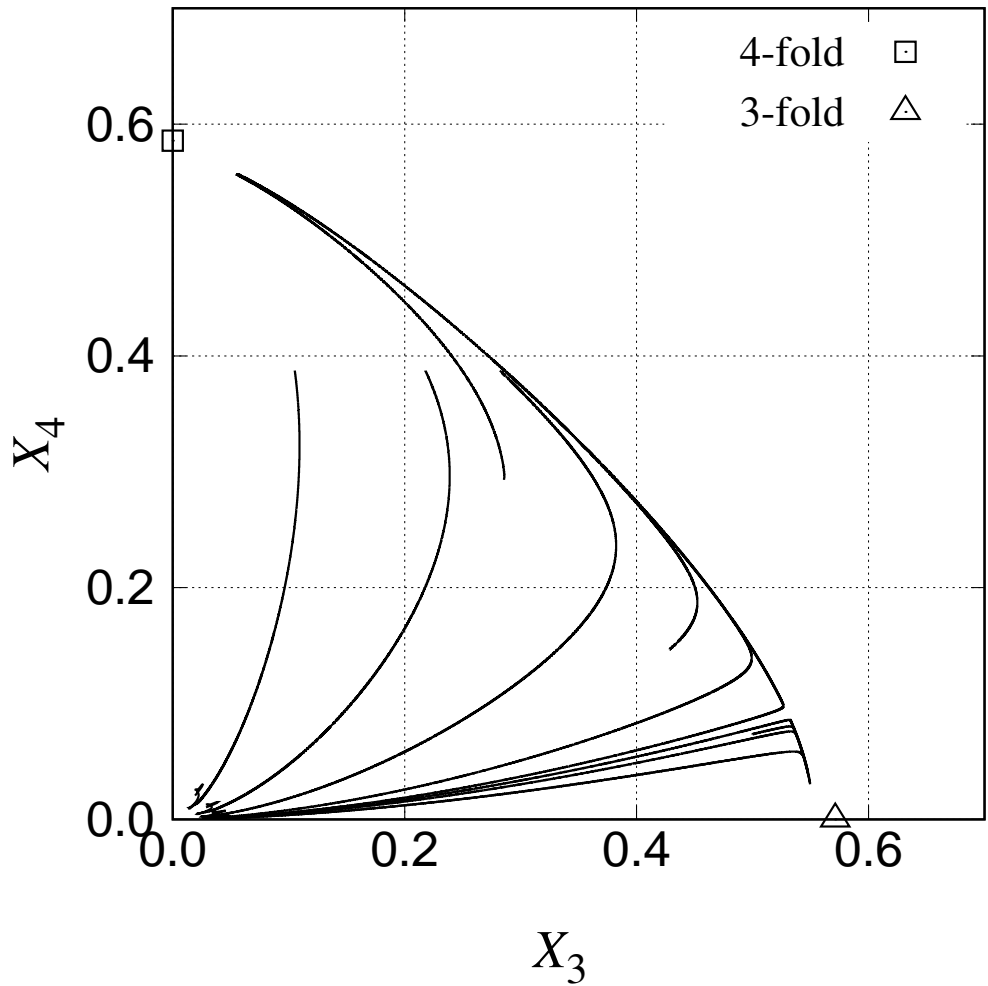}
  }
  \caption{
    Time variation of the system at $\Rey=490$ projected in the reduced space spanned by $X_3$ and $X_4$.
    The square on the ordinate indicates the $4$-fold spiral state bifurcated at $\Rey=\Rey_4$ from the axisymmetric state on the origin. The triangle on the abscissa indicates the $3$-fold spiral state performed at $\Rey=\Rey_3$.
  }
  \label{fig:m3-m4_Re=490}
\end{figure}

\begin{figure}
  \centerline{
    \includegraphics[angle=0,width=1.25\columnwidth]{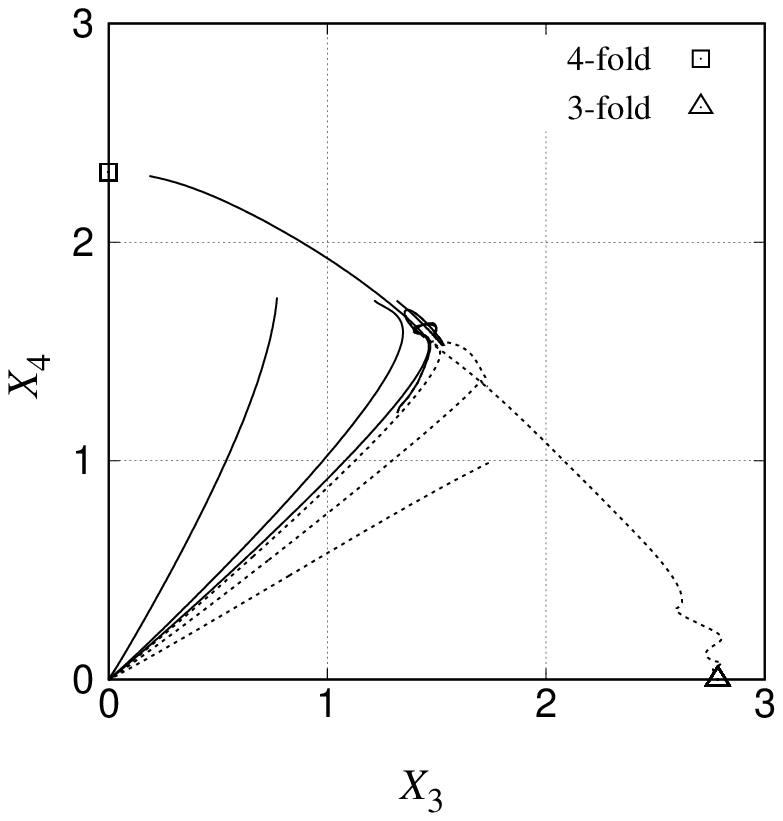}
  }
  \caption{
    Time variation of the system at $\Rey=560$ projected in the reduced space spanned by $X_3$ and $X_4$.
    Time variation of the states starting at different initial conditions are indicated as curves.
    The trajectories indicated by the solid and dashed curves approach the $4$- and $3$-fold spiral states, respectively.
  }
  \label{fig:m3-m4_Re=560}
\end{figure}

\SEC{$3$-fold spiral state coupled with unstable equilibria}
Trajectories on the projection obtained at $\Rey=490$, which is close to the first transitional Reynolds number $\Rey_4$ at $\eta=1/2$, are shown in Fig. \ref{fig:m3-m4_Re=490}. The time variation of the states starting at different initial conditions are indicated as curves, which separate either to the $3$- or $4$-fold spiral states at a saddle point on the hetero-clinic orbit.
Note again that the 4-fold spiral state emerged at a smaller Reynolds number than the 3-fold spiral state, that is, $\Rey_4<\Rey_3$.
The basin boundary emerged as a hidden concave curve with a small slope, which is identified in the figure as a watershed among trajectories starting near the origin and finally approaching either one of the spiral states in the projection.
It should also be noted that this basin boundary was closer to the abscissa than that obtained at $\Rey=560$, as shown in Fig. \ref{fig:m3-m4_Re=560}.
The $3$-fold spiral state at $\eta=1/2$, which bifurcates at a slightly higher Reynolds number than the $4$-fold spiral state, has a relatively narrow basin of attraction compared to that of the 4-fold spiral state at $\Rey=490$.
This hints that the $3$-fold spiral state and an unstable equilibrium bifurcate simultaneously from the origin at $\Rey=\Rey_3$, where the $3$-fold spiral state would be unstable otherwise.
The $4$-fold spiral state is the only stable equilibrium state at $\Rey_4<\Rey<\Rey_3$.
When the $3$-fold spiral state bifurcates at $\Rey=\Rey_3$, a boundary is formed between the basin of attraction of the $3$- and $4$-fold spiral states from the unstable equilibrium.
The trajectories concentrate on an arc connecting both stable states and they separate either to the $3$-fold or $4$-fold spiral states.
This implies that the arc is the projection of a hetero-clinic orbit connecting the two stable spiral states and that another hetero-clinic orbit connecting the basic and unstable equilibrium states as a saddle point on the basin boundary constitutes the basin boundary.

\SEC{B.B. at $\Rey=560$}
The trajectories on the projection obtained at $\Rey=560$ are shown in Fig. \ref{fig:m3-m4_Re=560}.
As the Reynolds number increased from the second transitional Reynolds number, the basin of the $3$-fold spiral state expanded in the state space and then became comparable with that of the $4$-fold spiral state.
From a deterministic viewpoint, the initial state, which is a point in the full state space, determined the final equilibrium that the state asymptotically reached.
In most experiments, the initial condition is uncontrollable such that the initial point can be selected randomly in the state space. Therefore, the final equilibrium state may be estimated by the ratio of the volume of the basin boundary in the state space.
The distance between the stable equilibrium state and unstable saddle point on the basin boundary in the full state space might be an index of the degree of attraction in the present bistable system.
The reproducibility of the final asymptotic state can be constructed using the index.

\begin{figure}
  \centerline{
    \includegraphics[angle=0,width=1.25\columnwidth]{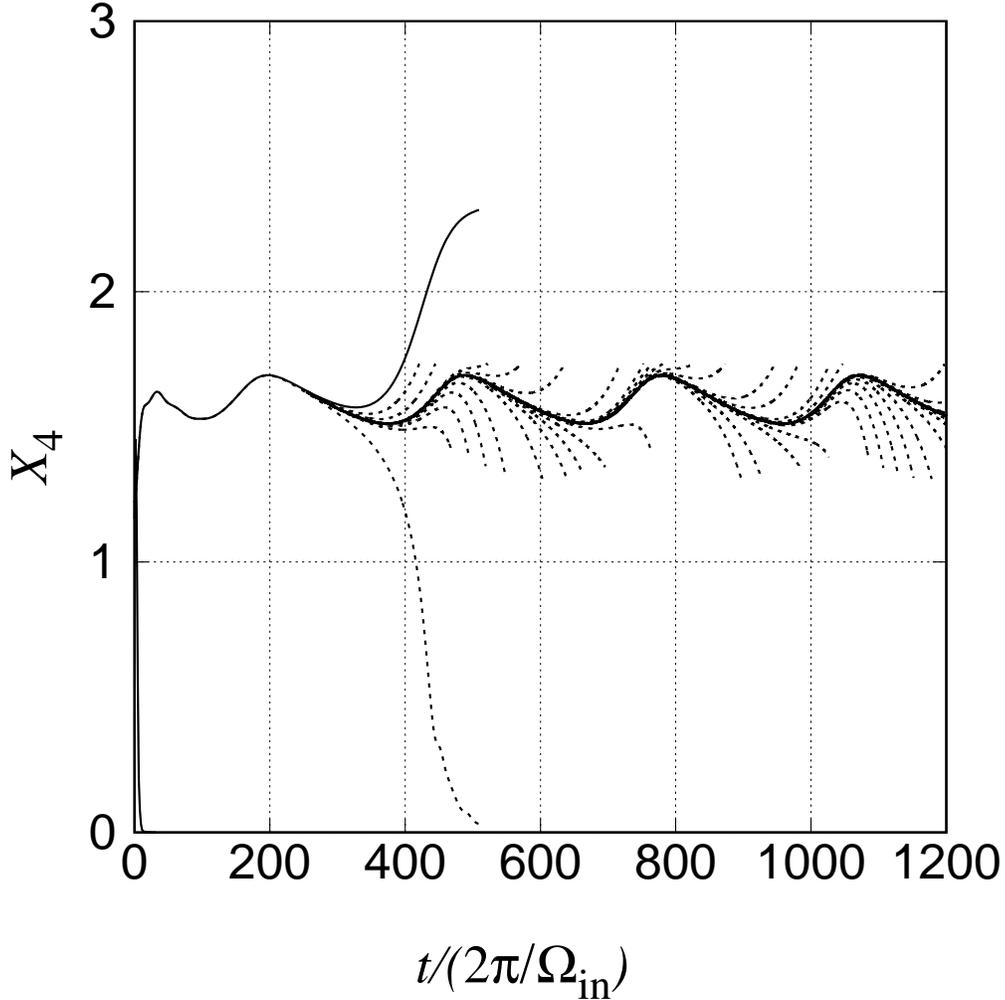}}
  \caption{Time variation of the degree of the 4-fold spiral state, $X_4(t)$, obtained through the edge-tracking at $\Rey=560$. Solid curves represent the trajectories starting from an intermediate state $\bfv{X}_{s}$ for $s<s_0$ approaching a 3-fold spiral state. Dotted curves represent trajectories for $s>s_0$ for a 4-fold spiral state.
  }
  \label{shooting_t}
\end{figure}

\SEC{Saddle on the BB at $\Rey=560$ is not a point}
The time variation of $X_4(t)$ for the trajectories obtained from the edge-tacking at $\Rey=560$ are shown in Fig. \ref{shooting_t}.
The unstable equilibrium state on the basin boundary was not a point but a periodic-like state composed of time-varying $X_3$ and $X_4$ components.
The obtained value of $s_0$ was insignificant because it depended on both $\bfv{X}_3$ and $\bfv{X}_4$, which were artificially adopted for the initial condition.
On the other hand, the achieved equilibrium state, independent from the initial condition, is significant and characterizes the system at a given $\Rey$.
Based on the visualization of this state, it was observed that the number of spiral arms extending from the poles to the equatorial zone in each hemisphere varied with time.
If the two spiral states with slightly different angular phase velocities at $\Rey=560$ would be superposed, the system would exhibit an interference in the time variation, observed as a beat between two distinct sounds with slightly different frequencies in terms of acoustics.
The values of $\omega_\phi/\Omega_{\rm in}$ were $0.1516$ and $0.1503$ for the $3$- and $4$-fold spiral states, respectively. Therefore, the period of the beat was $(2\pi/4/3)/\Delta \omega_\phi=403$, which is comparable to the period observed in Fig. \ref{shooting_t}.




\section{Summary}
  The bifurcation aspect of a wide-gap SCF was numerically investigated, with an emphasis on the competition among polygonal coherence with various wave numbers observed over the transitional Reynolds numbers.
  We focused on a representative case, the half-radius ratio $\eta=1/2$, by means of the continuation method based on the Newton--Raphson algorithm. This was used to confirm that the axisymmetric state became unstable over the first transitional Reynolds number $\Rey_4$ at which the 4-fold spiral state bifurcated.
  It was found that the 3-fold spiral state successively bifurcated from the axisymmetric state at a slightly higher Reynolds number $\Rey_3$ than the first transitional Reynolds number.
  The attraction of the 3-fold spiral state was guaranteed by a basin boundary consisting of a hidden unstable periodic-like state that bifurcates from the axisymmetric state at $\Rey_3$, which forms a hetero-clinic orbit connecting to the 3- and 4-fold spiral states in the state space.
  The attraction of the 3-fold spiral state expanded rapidly with an increase in the Reynolds number, which was verified using the distance from the unstable periodic-like state to both the spiral states in the state space.
  This aspect of the state space explains the experimentally bistable realization of different equilibrium states over the first transitional Reynolds number.
  It was also found that the periodic-like state was composed of the $3$- and $4$-fold spiral states, similar to a beat with two different frequencies.
  The validation of the present scenario under the other representative aspect ratio of the wide-gap SCF, such as $\eta=2/3$ and $3/4$, is ongoing.
  Future studies should attempt to understand the roles that a combination of a few polygonal modes plays along the route to turbulence in the wide-gap SCF under the Ruelle-Takens-Newhouse scenario.
  
\acknowledgments 
The authors would like to thank Dr. Yamashita, Dr. Yokoyama, Prof. J. Seki, and Prof. N. Sugimoto for their valuable comments.
We would also like to thank Editage (www.editage.com) for English language editing.
This work has been supported in part by KAKENHI (20K04294) and the European Union Horizon 2020 Research Innovation and Staff Exchange (RISE) program ATM2BT, grant number 824022, which includes Kansai and Akita Universities.

\bibliography{scf11}
\bibliographystyle{unsrt}

\end{document}